\documentclass[twocolumn]{aastex631}
\usepackage{isotope}
\usepackage{hyperref}

\usepackage{soul} 
\submitjournal{ApJ}

\shorttitle{Constraining nucleosynthesis in neutrino--driven winds}
\shortauthors{Psaltis et al.}
\graphicspath{{./}{figures/}}

\begin{document}

\title{Constraining nucleosynthesis in neutrino--driven winds: observations, simulations and nuclear physics}

\author[0000-0003-2197-0797]{A. Psaltis}
\affiliation{Institut f\"ur Kernphysik, Technische Universit\"at Darmstadt, 
Schlossgartenstr. 2, Darmstadt 64289, Germany}
\email{psaltis@theorie.ikp.physik.tu-darmstadt.de}

\author[0000-0002-6995-3032]{A. Arcones}
\affiliation{Institut f\"ur Kernphysik, Technische Universit\"at Darmstadt, 
Schlossgartenstr. 2, Darmstadt 64289, Germany}
\affiliation{GSI Helmholtzzentrum f\"ur Schwerionenforschung GmbH, Planckstr. 1, Darmstadt 64291, Germany}

\author{F. Montes}
\affiliation{National Superconducting Cyclotron Laboratory, East Lansing, MI, 48824, USA}
\affiliation{Joint Institute for Nuclear Astrophysics – CEE, Michigan State
University, East Lansing, Michigan 48824, USA}

\author[0000-0002-6695-9359]{P. Mohr}
\affiliation{Institute for Nuclear Research (ATOMKI), H-4001 Debrecen, Bem t\'er 18/c, Hungary}

\author[0000-0002-7277-7922]{C. J. Hansen}
\affiliation{Institute for Applied Physics, Goethe University Frankfurt, Max-von-Laue-Str. 12, Frankfurt am Main 60438, Germany}

\author{M. Jacobi}
\affiliation{Institut f\"ur Kernphysik, Technische Universit\"at Darmstadt, 
Schlossgartenstr. 2, Darmstadt 64289, Germany}

\author[0000-0003-1674-4859]{H. Schatz}
\affiliation{National Superconducting Cyclotron Laboratory, East Lansing, MI, 48824, USA}
\affiliation{Joint Institute for Nuclear Astrophysics – CEE, Michigan State
University, East Lansing, Michigan 48824, USA}
\affiliation{Department of Physics and Astronomy, Michigan State University, 567 Wilson Road, East Lansing, MI 48824, USA}



\begin{abstract}

A promising astrophysical site to produce the lighter heavy elements  
of the first $r$--process peak ($Z = 38-47$) is the moderately neutron rich ($0.4 < Y_e < 0.5$) neutrino--driven
ejecta of explosive environments, such as core--collapse supernovae and neutron star mergers, where the weak \textit{r}--process operates. This nucleosynthesis exhibits uncertainties from the absence of experimental data from $(\alpha,xn)$ reactions on neutron--rich nuclei, which are currently based on statistical model 
estimates. In this work, we report on a new study of the nuclear reaction impact using a Monte
Carlo approach and improved $(\alpha,xn)$ rates based on the Atomki-V2 $\alpha$ Optical Model Potential ($\alpha$OMP). We compare our results with observations from an up--to--date list of metal--poor stars with [Fe/H] $<$ -1.5 to find conditions of the neutrino--driven wind where the lighter heavy elements can be synthesized. 
We identified a list of $(\alpha,xn)$ reaction rates
that affect key elemental ratios in different astrophysical conditions. Our study aims on motivating more nuclear physics experiments on $(\alpha, xn)$ reactions using current and the new generation of radioactive beam facilities and also more observational studies of metal--poor stars.

\end{abstract}

\keywords{Core--collapse Supernova (304) --- Nuclear Astrophysics (1129) --- Nucleosynthesis (1131) --- R--process (1324) --- Nuclear reaction cross sections (2087)}


\section{Introduction} \label{sec:intro}

Solving the mystery of the origin of the heavy elements ($Z > 26$) in the universe has been a long--standing effort in nuclear astrophysics. Roughly half of them are produced via the rapid neutron capture process (\textit{r}--process)~\citep[for two recent reviews]{horowitz2019r, cowan2021origin}, although its astrophysical site or sites are still under discussion. The recent detection of a binary neutron star merger (NSM) via both gravitational waves (GW170817)~\citep{abbott2017gw170817} and the electromagnetic follow--up transient (AT2017gfo)~\citep{drout2017light} has reignited the interest for the origin of the \textit{r}--process. \cite{watson2019identification} identified strontium ($Z = 38$) in the merger ejecta, supporting the NSM as a site for \textit{r}--process nucleosynthesis. However, Galactic Chemical Evolution (GCE) modelling suggests that this cannot be the sole site~\citep{cote2019neutron} and other similarly exotic environments, such as Magneto--rotational Supernovae explosions (MR-SNe) can contribute to the Galactic \textit{r}--process abundance budget~\citep{Winteler:2012, Nishimura2017, reichert2021nucleosynthesis}.

Despite the ongoing discussions about its origin, the \textit{r}--process shows a unique robustness in its abundance pattern, with a couple of exceptions; the lighter region of the first peak, namely at $Z = 38-47$~\citep{sneden2008neutron} and also the actinides, where ``actinide--boost'' stars with enhanced thorium and uranium abundances 
compared to the main $r$--process have been recently observed~\citep{eichler2019probing, 2019ApJ...870...23H}. In addition to the $s$--process and $r$--process, the lighter heavy elements are also produced by an additional process (\textit{e.g.}, a light element primary process (LEPP)~\cite{travaglio2004galactic} and the weak $r$--process~\cite{montes2007}).

In recent years, there has been an intense observational effort to 
identify the elemental composition of metal--poor (old) stars, which are thought to be polluted by few or even
a single $r$--process event.~\cite{2004ApJ...607..474H, 2010ApJ...711..573R, Hansen2012, 2014AJ....147..136R, 2015ApJ...813...56N, 2017ApJ...837....8A},
to name a few, have focused on stars that show an enhancement in their lighter heavy elements compared to the
Z = 55--75 region of the solar $r$--process abundance pattern~\citep[Figure 11]{sneden2008neutron}. Such stars are called limited--$r$ or ``Honda--like'' stars, due to the seminal work of~\cite{2004ApJ...607..474H} in the giant HD 122563.
They are identified according to 
the standard classification, [Eu/Fe]\footnote{In the bracket notation [X/Y] = $\mathrm{\log \left( \frac{N(X)}{N(Y)}\right)_\star - \log \left( \frac{N(X)}{N(Y)}\right)_\odot}$, where the symbols $\star, \odot$ represent the stellar and solar values, respectively.} $<$ 0.3, [Sr/Ba] $>$ 0.5, and [Sr/Eu] $>$ 0.0~\citep{frebel2018nuclei, hansen2018r}. Metal--poor stars
that show a robust $r$--process pattern and have [Eu/Fe] $>$ +1.0 and [Ba/Eu] $<$ 0.0 are called $r$-II or ``Sneden--like'' stars, from the work of~\cite{2003ApJ...591..936S} in CS 22892-052 (also known as Sneden's star). This observational effort has offered valuable data to compare our nucleosynthesis theories to.

One of the proposed sites to produce the lighter heavy elements ($Z = 38-47$) in a primary process
is the slightly neutron--rich (0.4$< Y_e <$0.5) neutrino--driven ejecta of 
core--collapse supernovae explosions (CCSNe) or NSMs. In such conditions, a weak
\textit{r}--process (also referred to as $\alpha$--process) can 
operate~\citep{woosley1992alpha, witti1994nucleosynthesis, qian1996nucleosynthesis, hoffman1997nucleosynthesis, wanajo2001r, arcones2011production, hansen2014many, bliss2017impact, bliss2018survey}. 
The nucleosynthesis starts from Nuclear Statistical Equilibrium (NSE), since the ejected 
material is at high temperature and density, 
and when the temperature falls to $T_9 \approx 5$ -- where
$T_9$ is the temperature in units of $10^9$~K -- an $\alpha$--rich freeze--out
occurs. After that, the nucleosynthesis proceeds mainly through 
$\alpha$-- and proton--induced reactions on neutron--rich nuclei,
until the temperature drops to $T_9 \approx 2$. 
\cite{pereira2016theoretical} and \cite{bliss2017impact} have shown that the nucleosynthesis flow
strongly depends on $(\alpha,n)$ reactions on neutron--rich nuclei, which help to synthesize nuclei with larger atomic number $Z$.

Due to scarce experimental data, current weak \textit{r}--process nucleosynthesis
calculations employ $(\alpha,n)$ reaction rates based on the statistical 
Hauser--Feshbach formalism~\citep{hauser1952inelastic}. Unfortunately, such reaction rates can be uncertain as much as two orders of magnitude in the relevant temperature region ($T_9 = 2-5$)~\citep{pereira2016theoretical}. The
main contribution in the aforementioned uncertainties originates from the $\alpha$--nucleus potential ($\alpha$OMP), which has been identified as the prime nuclear physics uncertainty for this scenario~\citep{pereira2016theoretical, mohr2016role}.

Recently,~\cite{bliss2020nuclear} performed a nuclear reaction sensitivity study to explore
the impact of the $(\alpha,n)$ reactions for the weak \textit{r}--process nucleosynthesis using
reaction rates calculated from the Hauser--Feshbach code \texttt{TALYSv1.6}~\citep{koning2007talys}. In that
study the authors used the \texttt{TALYS} Global $\alpha$--Optical model Potential (GAOP) which 
is based on a spherical potential approach 
by~\cite{watanabe1958high}.~\cite{bliss2020nuclear} used a Monte Carlo technique to vary
all the $(\alpha,n)$ reaction rates in the network by random factors, sampled from a log--normal 
distribution with $\mu=0$ and $\sigma=2.4$, which corresponds to factors between 0.1 and 10 
in the 68.3\% coverage ($\log(10)= 2.4$). They identified a list of 45 ($\alpha,n$) 
reactions that impact the elemental abundances of the lighter heavy elements.

That work motivated numerous experimental studies, and new measurements of few $(\alpha,n)$ reaction cross sections for nuclei close to stability have been recently reported~\citep{kiss2021low, 2021PhRvC.104c5804S}. More experiments are currently proposed or 
are analyzed in nuclear physics facilities around the world.

In the present work, we build on the technique of~\cite{bliss2020nuclear}, using new, constrained rates of $(\alpha,xn)$ reactions, based on the Atomki-V2 $\alpha$OMP~\citep{mohr2020successful}. In addition, we
compare for the first time our nuclear reaction impact study results with abundance observations of metal--poor stars with [Fe/H]$<$-1.5 that show an enhancement in their lighter heavy element distribution and are considered candidates for the weak \textit{r}--process. For the astrophysical conditions that can reproduce the abundance observations, we identify a list of 21 $(\alpha, n)$ reactions that their rates need to be constrained experimentally.

This paper is structured as follows: in Section~\ref{sec:astro} we present the
different astrophysical conditions that we selected for our study. In Section~\ref{sec:compilation}, we present our up--to--date compilation of elemental abundances of metal--poor stars with an enhanced production of the first $r$--process elements. In Section~\ref{sec:atomki-v2}, we provide an introduction to the Atomki-V2 potential and its advantages. In Sections~\ref{sec:sensi} and~\ref{sec:results} we present the impact study and its results, along with the comparison to observational data. Finally, in Section~\ref{sec:conclusions} we conclude and discuss our results.

\section{Selection of the astrophysical conditions} \label{sec:astro}

\begin{figure}[hbpt!]
    \centering
    \includegraphics[width=.5\textwidth]{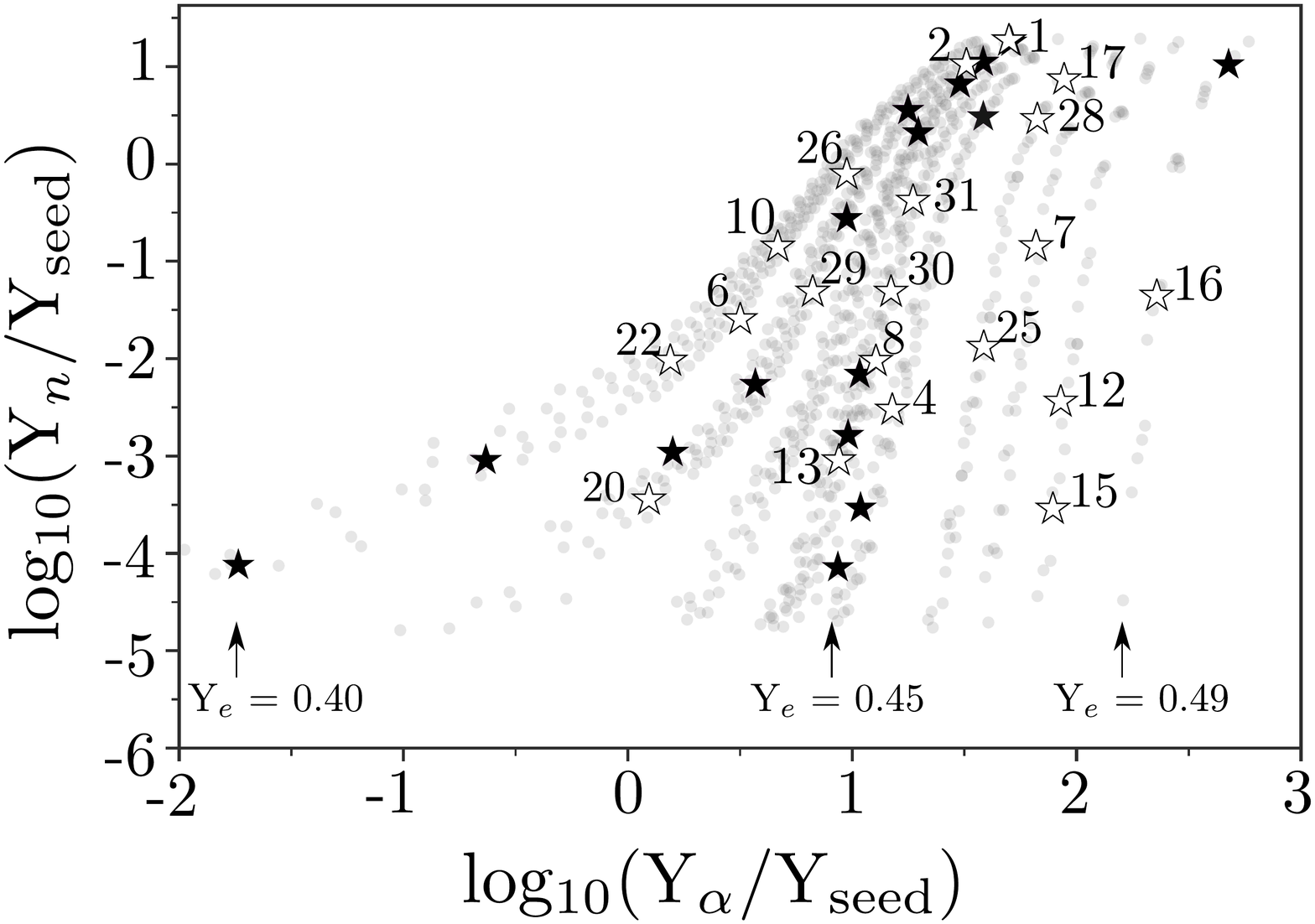}
    \caption{Distribution of the CPR2 tracers from~\cite{bliss2018survey} in the $\mathrm{Y_n/Y_{seed}}$--$\mathrm{Y_\alpha/Y_{seed}}$ phase space at $T_9 = 3$. The black stars represent the tracers studied in~\cite{bliss2020nuclear}. The subset of white stars were selected for this study. The arrows indicate the loci of tracers with the same $\mathrm{Y_e}$, which decreases with decreasing $\mathrm{Y_\alpha/Y_{seed}}$. 
    }
\label{fig:1}
\end{figure}

The conditions of the neutrino--driven wind ejecta are uncertain but also critical for the weak \textit{r}--process.~\cite{bliss2018survey} explored the relevant phase space ($Y_e$, entropy per baryon and expansion timescale) using a steady--state model. In the subsequent impact study of~\cite{bliss2020nuclear}, 36 representative trajectories from the CPR2 group -- the conditions that produce lighter heavy elements -- were chosen to test the importance of individual $(\alpha,n)$ reaction rates~\citep{nuc-astro}. In the present work, we study more than half of the aforementioned trajectories (20), which we summarize in Table~\ref{tab:1}. In Figure~\ref{fig:1} we map them in the neutron--to--seed ($\mathrm{Y_n/Y_{seed}}$)-- $\alpha$--to--seed ($\mathrm{Y_\alpha/Y_{seed}}$) space at $T_9 = 3$, where $\mathrm{Y_{seed}}$ is the sum of the abundances of all nuclei heavier than helium. Note that each loci of thermodynamic trajectories roughly corresponds to a different $Y_e$ of the wind ejecta.

\begin{deluxetable}{cccc}
\tablecaption{Main astrophysical conditions for each of the trajectories used in the present study. \label{tab:1}}
\tablehead{\colhead{Trajectory} & \colhead{$Y_e$} & \colhead{Entropy per baryon} &  \colhead{Expansion timescale}\\
& ~ & ~ ($k_B$/nucleon) & (ms)}
\startdata
MC01 & 0.42 & 129 & 11.7 \\
MC02 & 0.45 & 113 & 11.9 \\
MC04 & 0.44 & 66 & 19.2 \\
MC06 & 0.40 & 56 & 63.8  \\
MC07 & 0.47 & 96 & 11.6  \\
MC08 & 0.43 & 78 & 35.0  \\
MC10 & 0.40 & 54 & 31.0  \\
MC12 & 0.48 & 85 & 9.7   \\
MC13 & 0.43 & 64 & 35.9  \\
MC15 & 0.48 & 103 & 20.4 \\
MC16 & 0.49 & 126 & 15.4 \\
MC17 & 0.46 & 132 & 12.4 \\
MC20 & 0.41 & 42 & 59.3 \\
MC22 & 0.40 & 40 & 46.7  \\
MC25 & 0.46 & 96 & 20.9  \\
MC26 & 0.40 & 84 & 36.2  \\
MC28 & 0.46 & 113& 11.9  \\
MC29 & 0.41 & 66 & 41.4  \\
MC30 & 0.43 & 79 & 26.3  \\
MC31 & 0.43 & 71 & 11.4  \\
\enddata
\end{deluxetable}

\section{Compilation of abundance observations from metal--poor stars}
\label{sec:compilation}

In Table~\ref{tab:stars} we present an up--to--date list of elemental abundances from thirteen metal--poor stars with [Fe/H] $<$ -1.5 that are identified as ``Honda--like'' stars and also four ``Sneden--like'' stars. Typically metal--poor stars are those with [Fe/H] $<$ -1.0. However, the abundance cut we chose can eliminate stars that could have been polluted by Type Ia supernovae explosions. Most of the stars in our compilation have observed elemental abundances in the mass region of Z= 38-46, between strontium and palladium. Unfortunately, despite the wealth of observational data of metal--poor stars in the literature, one can find only a small sample of objects where more than few elements other than the Sr--Y--Zr triplet in the Z = 38--46 range are observed and reported~\citep{abohalima2018jinabase}. Studies that identify ``Honda--like'' stars,~\cite{hansen2018r} for example, only report measurements for select heavy elements, such as strontium, barium and europium. Although these elements are very important, since they are used for the usual classification of $r$--process stars, observations of niobium, molybdenum, ruthenium, palladium and silver are also crucial to help us identify the astrophysical conditions that produce the first $r$--process peak elements, as we discuss in Section~\ref{sec:conditions}.

\begin{deluxetable*}{lcccccccl}
\tablecaption{Observations of metal--poor stars used in the present work in units of $\log \epsilon$. The stars are grouped in ones that show a robust \textit{r}--process pattern (bottom -- ``Sneden--like'') and not (top -- ``Honda--like''). Typical uncertainties are $\delta X \sim 0.05-0.20$~dex. \label{tab:stars}}
\tablewidth{\linewidth}
\tablehead{\colhead{Star} & \colhead{Sr} & \colhead{Y} & \colhead{Zr} & \colhead{Nb} & \colhead{Mo} & \colhead{Ru} & \colhead{Pd} &  \colhead{Reference}}
\startdata
BD+42621 & 0.21(15) & -0.56(19) & 0.24(17) & \nodata & \nodata & \nodata &  \nodata &   \cite{Hansen2012} \\ 
BD+6648 & 0.95(19) &0.02(26) &0.76(18) & \nodata &0.03(25) &-0.31(28) &-0.78(21) & \cite{2017ApJ...837....8A} \\
CS 22942-035 & -0.28(26) &-1.45(18) &-0.41(18) &$<$0.45 &$<$-0.03 & \nodata & \nodata &\cite{2014AJ....147..136R} \\
HE 2217-0706 & 0.20(27)   & -0.68(22) & 0.03(22) & \nodata & \nodata & \nodata & \nodata &  \cite{barklem2005} \\
HD 4306 & -0.08(9) & -0.99(18) &-0.22(9) & \nodata & \nodata & \nodata & \nodata &\cite{2004ApJ...607..474H} \\ 
HD 23798 &0.86(19) &-0.04(26) &0.71(20) & \nodata &-0.11(25) &-0.17(28) &-0.74(21) &\cite{2017ApJ...837....8A} \\
HD 85773 & 0.00(20) &-0.96(28) &-0.23(19) & \nodata &-0.97(26) &-1.01(28) &-1.30(22)&\cite{2017ApJ...837....8A} \\ 
HD 88609 & -0.05(17)& -1.32(16)& -0.42(15)&$<$-0.33 &-1.31(36)& \nodata & \nodata &\cite{2014AJ....147..136R} \\
HD 107752 &-0.26(28) &-0.87(16) &-0.22(16) & \nodata &-0.90(17) &-0.96(21) &-1.35(18) &\cite{2017ApJ...837....8A} \\ 
HD 110184 &0.46(28) &-0.82(18) &-0.07(21) & \nodata &-0.70(17) &-0.85(21) &-1.22(18) &\cite{2017ApJ...837....8A} \\
HD 122563 & -0.12(14)&-0.93(20) &-0.28(15) &-1.48(14) &-0.87(19) &-0.86(20) &-1.88(21) &  \cite{2007ApJ...666.1189H} \\ 
HD 140283  & 0.24(33)& -1.00(22)&0.10(22) & \nodata & \nodata& \nodata & \nodata& \cite{2015ApJ...813...56N}  \\ 
HD 237846  &-0.35(26) &-1.60(18) &-0.69(17) &$<$-0.12 &$<$-0.63 & \nodata & \nodata&\cite{2010ApJ...711..573R} \\ 
\hline
CS 22892-052 & 0.45(13) & -0.42(10) &0.23(12) &-0.80(15) &-0.40(20) &0.08(10) & -0.29(10) & \cite{2003ApJ...591..936S} \\
CS 31082-001 &0.72(3) &-0.23(7) &0.43(15) &-0.55(20) &-0.37(22) &0.36(10) &-0.05(10) &\cite{Hill2002} \\
HD 115444 & 0.11(11) & -0.78(8) & -0.06(17) & \nodata & \nodata &-1.06(11) & \nodata &  \cite{2004ApJ...607..474H} \\
HD 221170 & 0.74(18) & -0.08(7) & 0.68(10) & -0.37(30)& 0.03(10)& 0.22(5)& -0.03(5) &\cite{2006ApJ...645..613I} \\
\enddata
\tablecomments{$\log \epsilon_X = \log \left( \frac{N_X}{N_H}\right) + 12$.}
\end{deluxetable*}

\section{A\lowercase{tomki-V2}: a new $\alpha$OMP}
\label{sec:atomki-v2}

As we mentioned in Section~\ref{sec:intro}, $(\alpha,n)$ reactions on 
intermediate and heavy mass nuclei, due to their high nuclear level density, are calculated 
within the statistical (Hauser--Feshbach) model. Within this framework, an $(\alpha,n)$ 
reaction cross section in the laboratory is defined as:
\begin{equation}
    \sigma_{(\alpha,n)} \sim \frac{T_{\alpha,0} T_n}{\sum_i T_i} = T_{\alpha,0} \cdot b_n
    \label{eq:statmod}
\end{equation}
where $T_i$ are the transmission coefficients for the i\textsuperscript{th} 
channel and $b_n = T_n/\sum_i T_i$ is the branching ratio for neutron decay.
Above the neutron threshold, the neutron transmission $T_n$ is dominating over the other 
transmissions $T_i$ and for this reason $b_n$ is close to unity. The compound nucleus 
formation cross section depends solely on $T_{\alpha,0}$ and thus only on the chosen $\alpha$OMP. It has been 
shown that different $\alpha$OMPs predict cross
sections that disagree up to two orders of magnitude at the astrophysically interesting 
energies \citep{pereira2016theoretical, mohr2016role}. This sensitivity mainly results from the tail of the imaginary part of the $\alpha$OMP at radii
outside the colliding nuclei \citep{mohr2020successful}. 

A different approach to calculate the compound nucleus formation is by employing a 
transmission through a real potential (pure barrier transmission model or PBTM). This approach avoids the above mentioned complications with the tail of the imaginary potential. The real part of the $\alpha$OMP is relatively well constrained; e.g., in the
case of the Atomki-V1 potential, the real part of the $\alpha$OMP is calculated from a double-folding approach, and its parameters are finetuned to experimental data of low--energy elastic
scattering~\citep{mohr2017statistical,mohr2013atomki-v1}. It was shown in \cite{mohr2020successful} that this approach is able to reproduce compound formation cross sections at very low sub--Coulomb energies with deviations below a factor of two over a wide range of masses. Recent experimental results~\citep{2021ApJ...908..202K,2021PhRvC.104c5804S} on stable isotope elements show a very good agreement with the predicted cross sections and thus confirm the predictive power of this approach.

The new Atomki-V2 $\alpha$OMP~\citep{mohr2020successful} combines the real part of the Atomki-V1 potential with a narrow, deep, and sharp-edged imaginary potential of Woods-Saxon type. The benefits of this combination are twofold. First, this parameterization ensures that the compound formation cross section is practically identical to the successful PBTM approach; thus, the excellent reproduction of the experimental compound formation cross sections persists. Second, the Atomki-V2 potential can be implemented in standard codes for nuclear reaction cross sections in the statistical model; thus, the branching ratios $b_i$ towards the different exit channels in Equation~(\ref{eq:statmod}) can be calculated in the usual way without further effort. 

A compilation of $\alpha$--induced reaction rates for nuclei between iron and bismuth ($26 \leq Z \leq 83$) using the Atomki-V2 $\alpha$OMP was recently published by~\cite{mohr2021astrophysical}. The calculations were performed using a modified version of the \texttt{TALYS} code~\citep{koning2007talys}. In general, the calculation of astrophysical reaction rates at high temperatures has to take into account that low--lying excited states in the target nuclei may be thermally populated. This compilation also includes the effect of thermal excitation and not just transmission from the ground state of the compound nucleus. For our impact study, we will use a conservative estimated uncertainty factor of three for the astrophysical reaction rates which is larger than a factor of two found between predicted and experimental cross sections of stable nuclei. This factor accounts for any increase in reaction rate uncertainties for nuclei away for stability, along the weak $r$--process path.

\section{Impact study on $\lowercase{(\alpha,n)}$ reaction rates}
\label{sec:sensi}

The impact study was performed using the nuclear reaction
network \texttt{WinNet}~\citep{Winteler:2011,Winteler:2012}
using 4053 nuclei up to hafnium ($Z= 72$), which are connected with
$\approx$ 54,000 reactions. Theoretical weak reactions are taken from \citet{Langanke:2001}, and 
neutrino reactions from \citep[see also \citealt{Froehlich:2006b} for details about the   
neutrino reactions]{Langanke:2001b}. All other reaction rates, except for the $(\alpha,n)$ reactions, are adopted from the JINA REACLIB 
compilation~\citep{Cyburt:2010}. For the calculations of the neutrino--driven 
trajectories from the study of~\cite{bliss2020nuclear}, the electron 
fraction $\mathrm{Y_e}$ evolves in nuclear statistical equilibrium (NSE) at a temperature of $T_9= 10$ and assume that NSE stands down to $T_9= 7$. 

For each selected trajectory, we performed $10^4$ network calculations varying 
simultaneously all $\sim$ 4300 $(\alpha,xn)$ reactions between iron (Z=26) 
and hafnium (Z= 72). Specifically, each $(\alpha,xn)$ reaction was multiplied by
a randomly sampled variation factor $p$. The sampling was performed from a 
log--normal distribution with $\mu=0$ and $\sigma=1.10$, which corresponds to factors
between 0.33 and 3 in the 68.3\% coverage. We selected these limits, due to the 
success of the Atomki-V2 $\alpha$OMP to agree with experimental data to within a factor of 3. Log--normal distributions are ideal for such studies because they are defined only for $p \geq 0$~\citep{parikh2008effects, longland2010charged, rauscher2016uncertainties, nishimura2019uncertainties, bliss2020nuclear}. The log--normal distribution is a versatile tool for statistical analysis that is used in a wide variety of disciplines, such as finance~\citep{black1991bond} and epidemiology~\citep{linton2020incubation}. Since the forward and reverse reactions are connected by detailed balance in weak $r$--process conditions, we used the same variation factor for the latter. One might argue that multiplying all the relevant reaction rates by a constant factor for the whole temperature range produces unrealistic results for the final abundances, since it omits any temperature dependence they might have. \cite{longland2012recommendations} showed that this approach is in fact a good approximation when using theoretical reaction rates. In particular, the temperature dependence of a rate has a minimal effect in nucleosynthesis yields, when many rates are sampled and varied simultaneously.

\section{Results} \label{sec:results}

\begin{figure}[ht!]
    \centering
    \includegraphics[width=.5\textwidth]{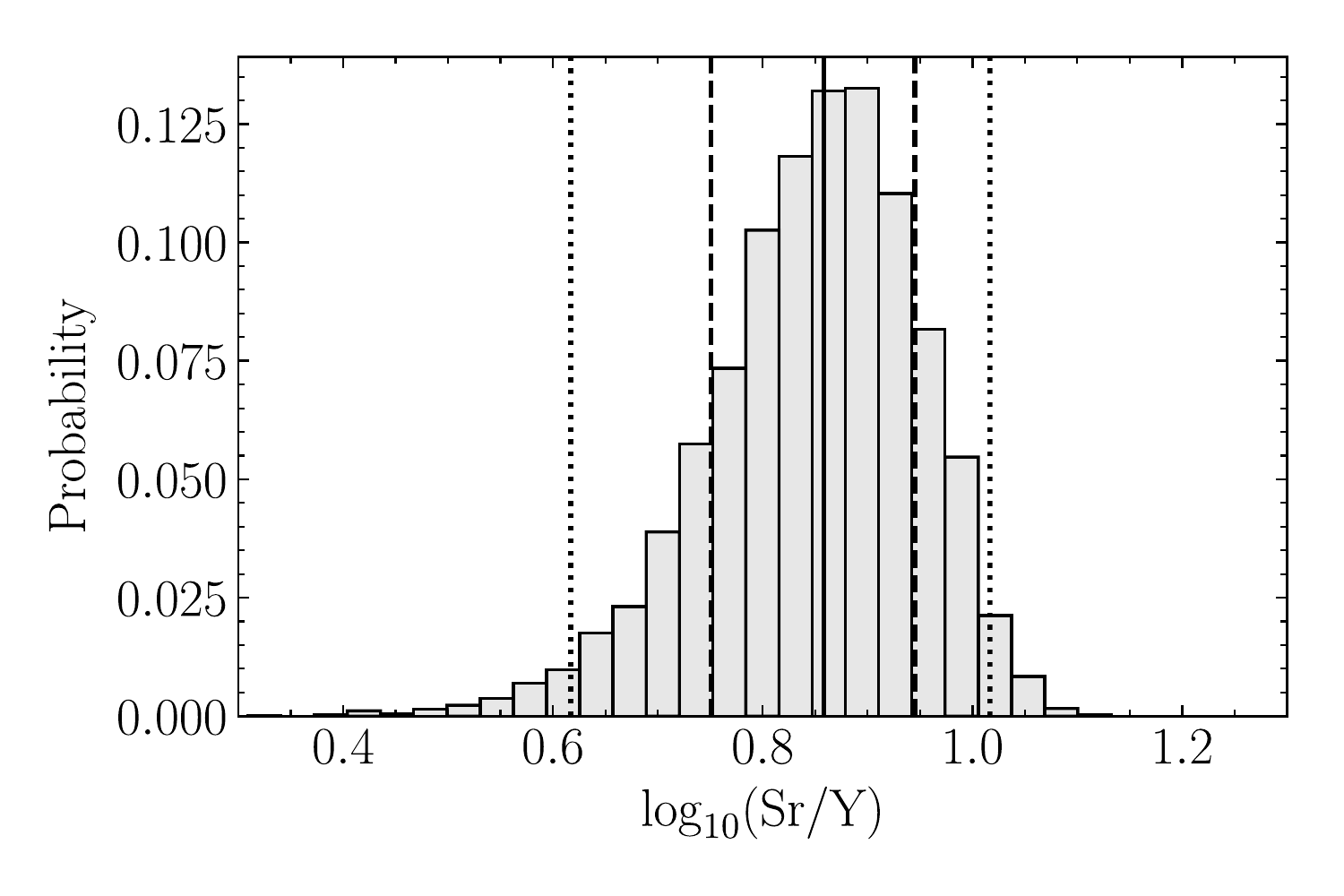}
    \caption{Probability histogram of the $\mathrm{\log_{10}(Sr/Y)}$ ratio from $10^4$ nucleosynthesis calculations of the trajectory \texttt{MC08}. The solid line shows the 50\textsuperscript{th} percentile while the dashed and dotted lines indicate the 1 and 2$\sigma$ uncertainties (68\% and 95\% confidence intervals), respectively.}
\label{fig:2}
\end{figure}

In this section we discuss the main results of our impact study and the comparison to observational data. Figure~\ref{fig:2} shows a probability histogram of the $\mathrm{\log_{10}(Sr/Y)}$ ratio, which is also one of the main observables in metal--poor stars. The strontium--yttrium--zirconium triplet is of extreme importance in nuclear astrophysics, since it can be produced by
a variety of processes~\citep{travaglio2004galactic}.
To select the number and widths of the bins in Figure~\ref{fig:2}, we use the Freedman--Diaconis 
rule~\citep{freedman1981histogram} which takes into account both the sample size, and its spread. 
We can create similar graphs to compare our impact study results with observations of metal--poor stars in a statistically meaningful manner, as we shall present in Section~\ref{subsec:observations}.

\begin{figure}[hbpt!]
    \centering
    \includegraphics[width=0.5\textwidth]{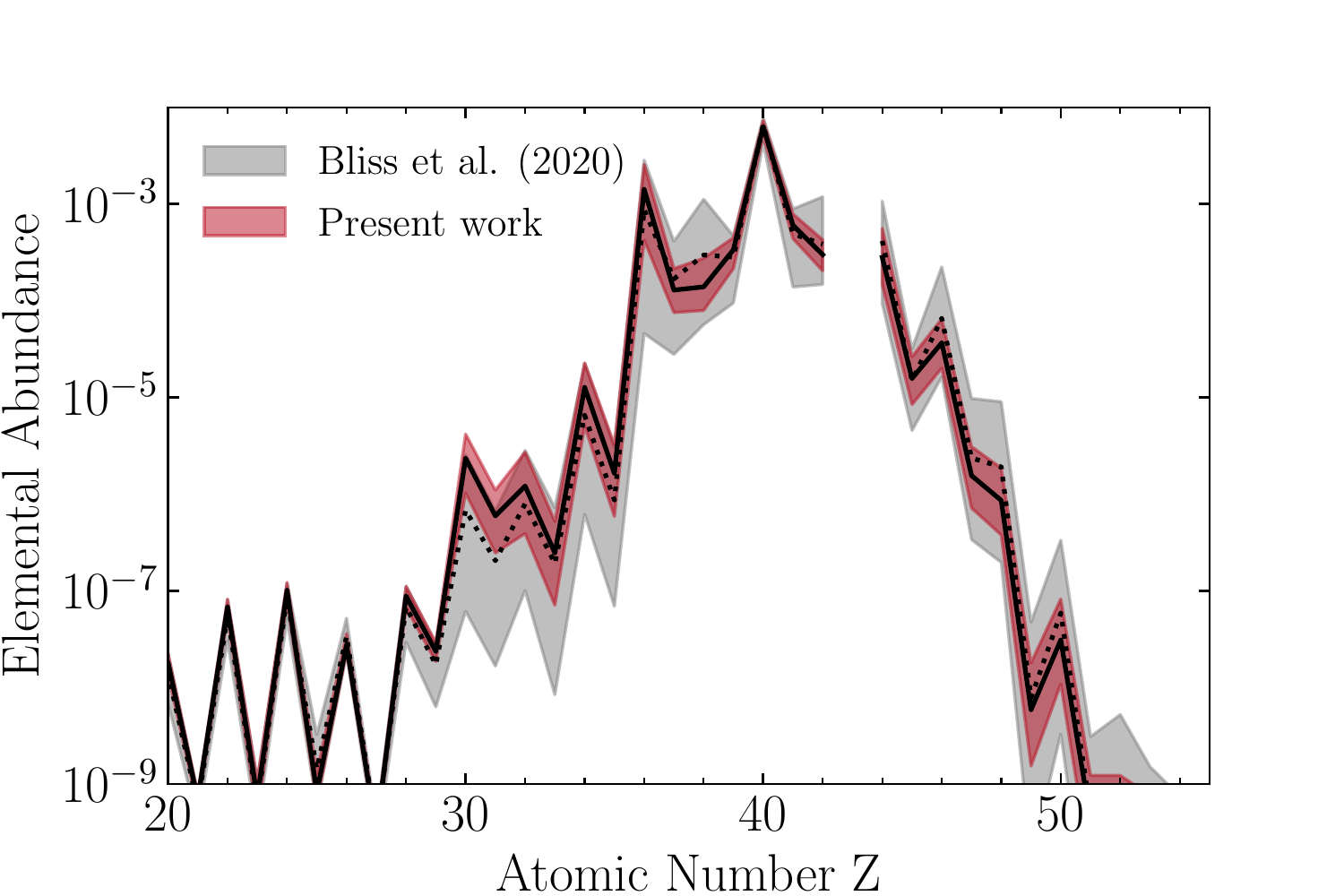}
    \caption{Comparison of the final abundances for trajectory \texttt{MC06} from the work of~\cite{bliss2020nuclear} (grey and dotted line) and the present work (red and solid line). In each case, the bands show the 2$\sigma$ uncertainties due to $(\alpha, n)$ reaction rates.}
\label{fig:3}
\end{figure}

We choose the representative trajectory \texttt{MC06} ($\mathrm{Y_e} = 0.40$, $s = 56$ $k_B$/nucleon, $\tau = 63.6$~ms) as a benchmark to compare our results with the work of~\cite{bliss2020nuclear}. Figure~\ref{fig:3} shows 
the elemental abundance distribution at time $t= 1$~Gy, when all nuclei in the network have decayed to stability. The two calculations agree well, and in addition, we can note that the overall uncertainty of our calculations is smaller, which can be attributed to the fact that we are using a factor of 3 lower $(\alpha,xn)$ reaction rate uncertainties.

\begin{figure*}[ht!]
    \centering
    \includegraphics[width=\textwidth]{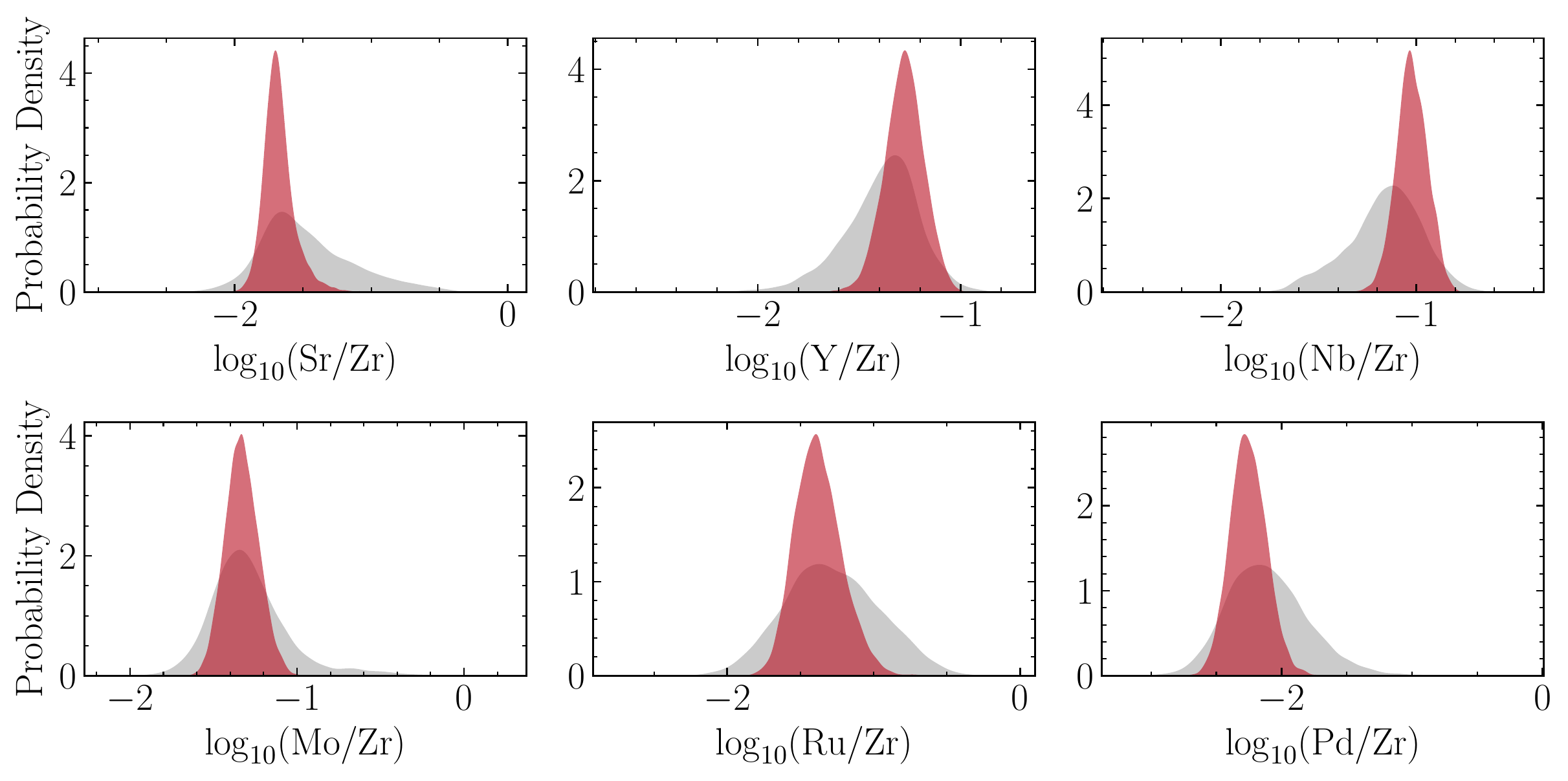}
    \caption{Comparison of the probability density KDE of six elemental abundance ratios between~\cite{bliss2020nuclear} (grey) and our present work (red) for the trajectory \texttt{MC06}. See the text for details.}
\label{fig:4}
\end{figure*}

\begin{figure}[hbpt!]
    \centering
    \includegraphics[width=.5\textwidth]{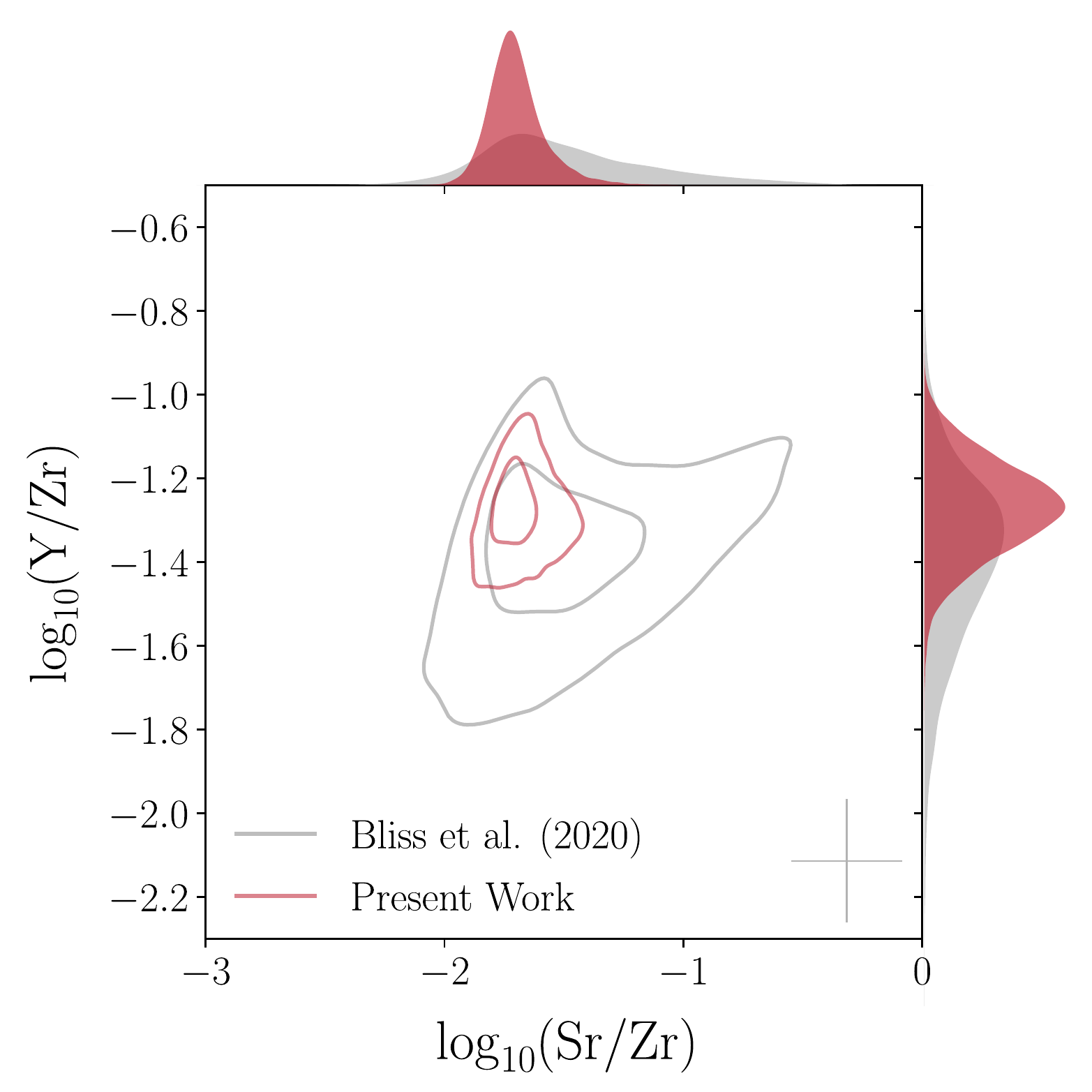}
    \caption{Bivariate probability density KDE for the $\mathrm{\log_{10}(Sr/Zr)}$ and 
    $\mathrm{\log_{10}(Y/Zr)}$ ratios from $10^4$ nucleosynthesis calculations of the trajectory \texttt{MC06}. The contours show the 1 and 2$\sigma$ confidence intervals of each distribution. The individual KDEs are shown in the margins. Typical observational uncertainties are shown in the bottom right.}
\label{fig:5}
\end{figure}

In Figure~\ref{fig:4} we compare the kernel density estimates (KDEs)~\citep{izenman1991review} for a select list of
elemental abundance ratio distributions between our study and~\cite{bliss2020nuclear}. Once again, the distributions using the constrained $(\alpha, xn)$ reaction rate uncertainties based on the Atomki-v2 $\alpha$OMP are much narrower compared to the GAOP ones, and in most cases the $1 \sigma$ uncertainty is comparable to observational errors~\citep[for example]{2004ApJ...607..474H} ($\delta X \sim 0.2$~dex). This can be further illustrated in Figure~\ref{fig:5}, where we show the bivariate KDE of $\mathrm{\log_{10}(Y/Zr)}$ versus $\mathrm{\log_{10}(Sr/Zr)}$ from the representative trajectory \texttt{MC06}. Note that the 1 and 2$\sigma$ contours cover the 39.3\% and 86.5\% of the total volume, respectively.

\subsection{Comparison to elemental abundance ratios of metal--poor stars}
\label{subsec:observations}

\begin{figure*}[hbpt!]
    \centering
    \includegraphics[width=\textwidth]{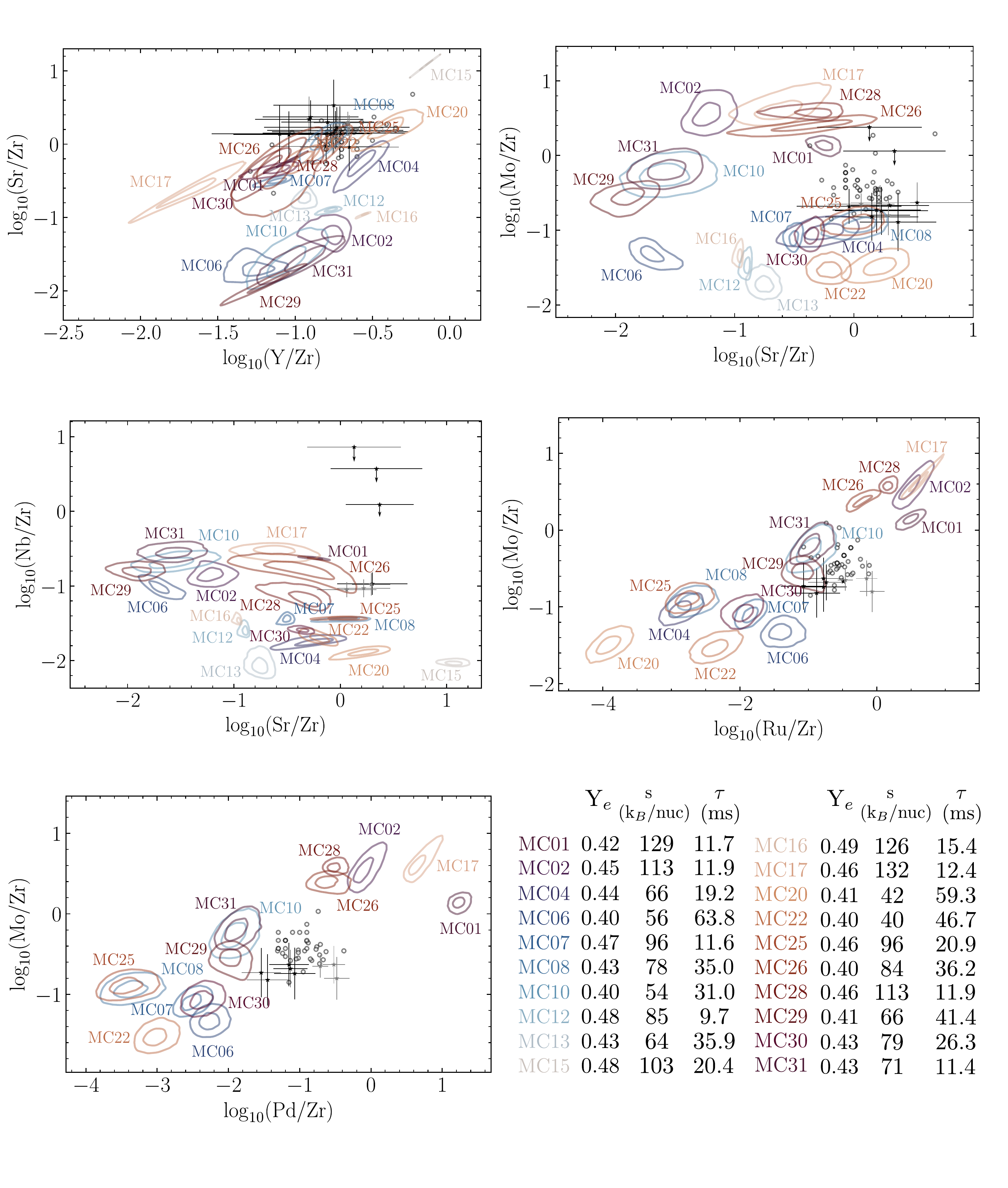}
    \caption{Bivariate KDEs for five elemental ratio pairs for the Monte Carlo trajectories studied in the present work. The distributions depict the 1 and 2$\sigma$ contours that cover the 39.3\% and 86.5\% of the total volume, respectively. Black and grey stars represent Honda--like ($r$--limited) and Sneden--like ($r$--II) stars from Table~\ref{tab:stars}, respectively. The open circles show additional observations from the works of~\cite{peterson2013} and~\cite{Hansen2012}.}
\label{fig:6} 
\end{figure*}

The novelty of the present impact study is that we focus on \textit{elemental abundance ratios} instead of single elemental abundances. In Figure~\ref{fig:6} we compare bivariate probability density KDEs for all the Monte Carlo trajectories to the observations of Table~\ref{tab:stars} for different combinations of elemental ratio pairs, namely Sr/Zr, Y/Zr,  Nb/Zr, Mo/Zr, Ru/Zr and Pd/Zr. Note that in some cases, for example the Pd/Zr vs Mo/Zr ratios not all MC trajectories are shown, since their production of Z $>$ 42 elements is below our abundance threshold of $Y_\mathrm{min} = 10^{-10}$. It should be noted that for all the ratios shown in Figure~\ref{fig:6}, no overproduction of other abundances is found. 

We chose the above elemental ratio pairs to investigate the nuclear physics 
impact via the $(\alpha, n)$ reaction rates to the production of elements with the most observational statistics, that is the Sr--Y--Zr triplet, the Z = 38--42 and the Z = 42--46 regions, respectively. In addition to the Table~\ref{tab:stars} compilation, we included additional data from the works of~\cite{peterson2013} and~\cite{Hansen2012}, to provide a larger sample of metal--poor stars, that are not necessarily identified as ``Honda--like'' or ``Sneden--like''.

As Figure~\ref{fig:6} demonstrates, using elemental abundance ratios instead of elemental abundances is a powerful tool to constrain the astrophysical conditions of the weak $r$--process. By reducing the associated nuclear physics uncertainties -- experimentally determining the relevant $(\alpha, xn)$ reaction rates (Section~\ref{sec:reactions}) to reduce the size of the contours -- and more precise observations (smaller error bars) we would be able to identify the conditions where the lighter heavy elements can be produced in the neutrino--driven wind ejecta.

\subsection{Which astrophysical conditions can produce the lighter heavy elements?}
\label{sec:conditions}

According to our analysis, we can identify individual conditions of the neutrino--driven ejecta where the lighter heavy elements can be produced. This is demonstrated in Figure~\ref{fig:6}, where bivariate KDE contours are compared with the elemental abundance ratios from the metal-poor stars compiled in Section~\ref{sec:compilation}. Figure~\ref{fig:6} depicts the interconnection between three different aspects of nuclear astrophysics: (i) elemental abundance observations in metal--poor stars (data points), (ii) astrophysical modeling (location of the different contour lines) and (iii) the nuclear physics impact (size are covered area of the contours).

\begin{figure}[hbpt!]
    \centering
    \includegraphics[width=0.45\textwidth]{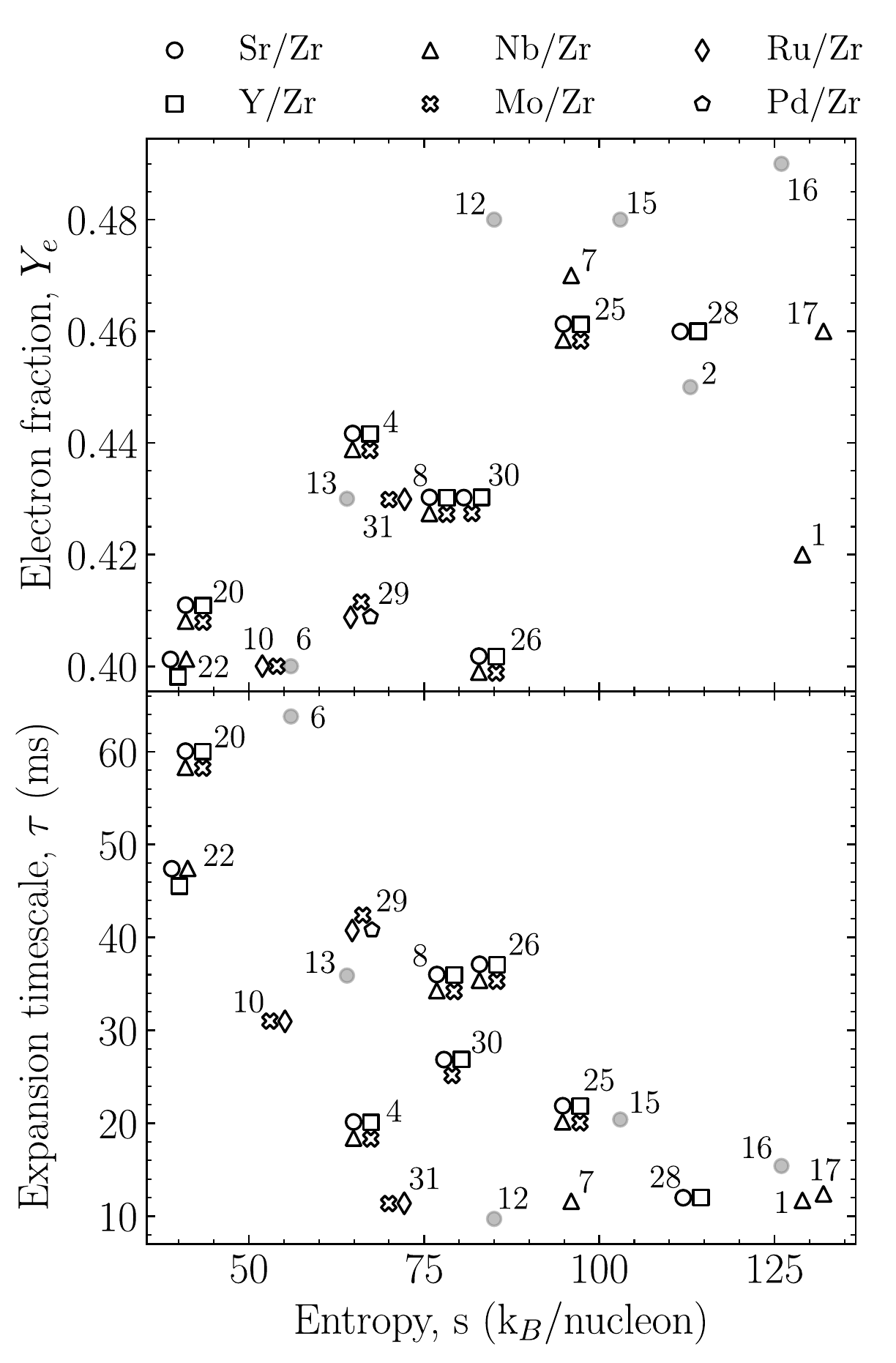}
    \caption{ (Top panel) Distribution of the MC trajectories in the entropy vs. $Y_e$ phase space. The different symbols represent the elemental ratios that match stellar observations for a given trajectory. The astrophysical conditions for each trajectory are given in Table~\ref{tab:1}. The grey points are MC trajectories used in the present work that do not reproduce any elemental ratio from our observational compilation. (Bottom panel) Similar to the top panel but for the entropy vs. expansion timescale phase space. See the text for details.}
\label{fig:7} 
\end{figure}

In Figure~\ref{fig:7} we map the thermodynamic trajectories that reproduce the observed abundance ratios of Figure~\ref{fig:6} in the entropy per bayon versus $Y_e$ and entropy per baryon versus expansion timescale phase spaces to gain some more insight about the astrophysical conditions that can produce the first $r$--process peak elements. In the upper panel, we observe that with the exception of tracers 25 and 28, the low entropy per baryon, $s \lesssim~85~ k_B$/nucleon, and $Y_e \lesssim 0.44$ are more likely to reproduce the observations. As it has been discussed in detail in the literature~\citep[for example]{woosley1992alpha, qian1996nucleosynthesis}, the entropy per baryon in neutrino--driven winds is related to temperature and density, $s \propto T^3/\rho$. High entropy per baryon leads to more free nucleons and less seed nuclei, which results in a larger neutron--to--seed ratio and thus an increased production of lighter heavy nuclei~\citep{arcones2014}. Note that in the case of the Nb/Zr ratio ($\triangle$ in Figure~\ref{fig:7}), given the scarcity of observational data (only four ``Honda--like'' stars in Table~\ref{tab:stars}) and the fact that most of the stars in our compilation have only observational upper limits, there are many conditions that can reproduce it within uncertainties.

In the lower panel of Figure~\ref{fig:7} we map the thermodynamic trajectories of our study in the entropy per baryon vs. expansion timescale space. The expansion timescale exhibits similar effects to the entropy, with fast winds (low expansion timescale $\tau$) leading to lower seed nuclei available for nucleosynthesis, compared to a slower wind (higher expansion timescale $\tau$).

Trying to identify the astrophysical conditions that produce specific regions of the Z= 38--46 region, we can find connections in the way that the entropy per baryon, $Y_e$ and expansion timescale affect the weak $r$--process. Specifically, for the Sr/Zr and Y/Zr ratios, for which we have the largest observational dataset, Figure~\ref{fig:7} shows that 
conditions with relatively low $Y_e$ ($0.40 \lesssim Y_e \lesssim 0.46$) and entropy per baryon ($ 55 \lesssim s \lesssim 85$ $k_B$/nucleon) are more favorable to match these observations. We should stress that the neutrino--driven ejected material, such as in a core--collapse supernova explosion~\citep{witt2021}, follows a distribution of Y$_e$, entropy per baryon, and expansion timescale, meaning that there is not a single astrophysical condition in a given explosion. Rather, it is the combination of many different conditions that provide the final abundance pattern.

Interestingly enough, there are only a handful of tracers that can reproduce the heavier elemental ratios, Ru/Zr and Pd/Zr (10, 29 and 31). These three conditions have entropy per baryon $ 55 \lesssim s \lesssim 75$ $k_B$/nucleon and $0.40 \lesssim Y_e \lesssim 0.43$. 

The works of \cite{hansen2014many} and~\cite{arcones2014} have explored a wide range of conditions of the neutrino--driven wind ejecta and focused on the production of Sr, Y, Zr, and Ag both in neutron-- and proton--rich winds. One of their main results, which we can confirm in our study, is that small variations in the conditions, produce large effects in the final abundances of the lighter heavy elements.

Another interesting result, shown in Figure~\ref{fig:6}, is that even though we use the same uncertainty for the $(\alpha, n)$ reaction rates for all the Monte Carlo trajectories, the size of the contours for different astrophysical conditions are different. This can be attributed to the sensitivity of each condition to changes in the main nuclear reaction channel, $(\alpha,n)$ on neutron---rich nuclei, which moves the nucleosynthesis flow to heavier species.

It is worth adding that proton--rich conditions of the neutrino--driven wind can also produce the elements between strontium and silver, via the $\nu p$--process ~\citep{Froehlich:2006b}. This scenario operates in the neutron--deficient side of the chart of nuclides and the nucleosynthesis is flowing via sequences of $(p, \gamma)$ and $(n, p)$ reactions. \cite{arcones2014} have showed that in the $\nu p$--process, the abundance pattern of Z= 38--47 elements is homogeneous and changes smoothly when varying the wind parameters (entropy, expansion timescale and $Y_e$), in contrast to the neutron--rich conditions.

\subsection{The most important $(\alpha,n)$ reactions}
\label{sec:reactions}

We have identified the most important reactions for each astrophysical condition of the neutrino--driven ejecta by the correlation between reaction rate variation and elemental abundance ratio. For this, we employ the Spearman's correlation coefficient $\mathrm{r_{corr}}$~\citep{spearman1904spearman}, which is defined as:
\begin{equation}
    r_{corr} = \frac{\sum_{i=1}^n (R(p_i)- \overline{R(p)}) (R(Y_{ri})-\overline{R(Y_r)})}{\sqrt{\sum_{i=1}^n (R(p_i)- \overline{R(p)})^2} \sqrt{\sum_{i=1}^n (R(Y_{ri})-\overline{R(Y_r)})^2}}
\end{equation}
where $n = 10^4$ is the number of calculations we performed for each thermodynamic trajectory, $R$ denotes the ranks of a rate variation $\{p_1, p_2, \cdots, p_n \}$ and the final abundance ratio $\{ Y_{r1}, Y_{r2}, \cdots Y_{rn} \}$. $\overline{R(p)}$ and $\overline{R(Y_r)}$ are the average ranks for rate variation and elemental abundance ratio. $\mathrm{r_{corr}}$ lies in the $[-1,1]$ space, and (-)1 shows a perfect monotonic (anti-)correlation of the two quantities.
The Spearman's correlation is suitable for such studies, due to the non--linearity
between reaction rate variations and resulted abundances. In previous nuclear reaction
sensitivity studies~\citep[for example]{nishimura2019uncertainties, rauscher2016uncertainties} the Pearson correlation coefficient~\citep{pearson1895correlation} has been employed, which instead evaluates the linear relationship between the two aforementioned quantities.

Using the above analysis, we investigated for the first time which $(\alpha, xn)$ reaction rates affect \textit{elemental abundance ratios} of the first $r$--process peak elements that are observed in metal--poor stars. We identified $(\alpha, xn)$ reactions as important when their $|\mathrm{r_{corr}}| > 0.20$ for a specific ratio. 
We investigated all the thermodynamic trajectories of Table~\ref{tab:1}, despite the fact that some of them do not match the observations in Figure~\ref{fig:6}. As we discussed in Section~\ref{sec:conditions}, a combination of different astrophysical conditions could also explain the observations in metal--poor stars. In the Appendix we also report our results for $(\alpha, n)$ reactions that affect elemental abundances for all the MC trajectories of Table~\ref{tab:1} and compare our results with the work of~\cite{bliss2020nuclear}.

To guide the experimental nuclear physics community for future measurements, we have grouped these 35 $(\alpha, xn)$ reactions rates, that are also listed in Table~\ref{tab:reactions_ratios}, according to how many ratios affect and in how many conditions of the neutrino--driven wind.

\begin{enumerate}
    \item[] \textit{Many elemental ratios under many astrophysical conditions}: $\isotope[84][]{Se}$, $\isotope[87-89][]{Kr}$, $\isotope[93][]{Sr}$, 
    \item[] \textit{Few elemental ratios under many astrophysical conditions}:
      $\isotope[86][]{Br}$, $\isotope[86, 90][]{Kr}$, $\isotope[87-89][]{Rb}$, $\isotope[91, 92, 94][]{Sr}$, $\isotope[94][]{Y}$
    \item[] \textit{Many elemental ratios under few astrophysical conditions}: 
     $\isotope[85][]{Se}$, $\isotope[85][]{Br}$,
    \item[] \textit{Few elemental ratios under few astrophysical conditions}: $\isotope[63][]{Co}$, $\isotope[67][]{Cu}$, $\isotope[79, 81][]{Ga}$ $\isotope[76][]{Zn}$, $\isotope[80,82][]{Ge}$, $\isotope[83][]{As}$, $\isotope[87, 90,91][]{Rb}$, $\isotope[88-90][]{Sr}$, $\isotope[95, 96][]{Y}$,
    $\isotope[96-98][]{Zr}$.
\end{enumerate}

\begin{deluxetable*}{cccc}
\tablecaption{$(\alpha, xn)$ reaction rates for which the Spearman's coefficient is $|\mathrm{r_{corr}}|\geq0.20$ in neutrino--driven wind trajectories for the given elemental ratios of Figure~\ref{fig:6}. Normal text entries indicate the ones that match the abundance ratios within 1$\sigma$ in observational errors and 2$\sigma$ confidence intervals based on the variation of $(\alpha, xn)$ reaction rates. The entries in \textit{italics} do not match the observed abundance ratios, but are included for completeness. See the text for details. \label{tab:reactions_ratios}}
\tablehead{\colhead{Reaction} & \colhead{Affected elemental ratios} & \colhead{Correlation Coefficient, $|\mathrm{r_{corr}}|$} &  \colhead{MC Tracers}}
\startdata
$\mathit{^{63}Co(\alpha,n)}$ & \textit{Sr/Zr, Y/Zr}   & \textit{0.20-0.29}  &  \textit{13, 15}   \\
$\mathit{^{67}Cu(\alpha,n)}$ &\textit{Sr/Zr, Y/Zr}   & \textit{0.20-0.39}  &  \textit{12, 13}   \\
$\isotope[76][]{Zn}(\alpha,n)$ & Sr/Zr, \textit{Y/Zr}, Nb/Zr, \textit{Mo/Zr, Ru/Zr, Pd/Zr}    & 0.29-0.37, \textit{0.25-0.33}  &  1   \\
$\mathit{^{79}Ga(\alpha,n)}$ & \textit{Mo/Zr, Ru/Zr, Pd/Zr} & \textit{0.25-0.33}  &  \textit{1}   \\
$\mathit{^{81}Ga(\alpha,n)}$ & \textit{Mo/Zr} & \textit{0.22}  &  \textit{1}   \\
$\isotope[80][]{Ge}(\alpha,n)$ & Sr/Zr, \textit{Y/Zr}, Nb/Zr, \textit{Pd/Zr}   & 0.23-0.35, \textit{0.33-0.34}  &  1, 17   \\
$\isotope[82][]{Ge}(\alpha,n)$ & Sr/Zr, \textit{Y/Zr}, Nb/Zr,  \textit{Ru/Zr}, \textit{Pd/Zr}  & 0.24-0.60, \textit{0.25-0.59}  &  1, 17   \\
$\isotope[83][]{As}(\alpha,n)$ & Sr/Zr, \textit{Mo/Zr}  & 0.49, \textit{0.55}  & \textit{1}, 17   \\
$\isotope[84][]{Se}(\alpha,n)$ & Sr/Zr, Y/Zr, Nb/Zr, Mo/Zr, \textit{Ru/Zr, Pd/Zr}  & 0.23-0.91, \textit{0.32-0.84}  &  \textit{1, 2, 6}, 7, \textit{10}, 17, 20   \\
~ & ~ & ~ & 22, 26, 28, 29, 30, \textit{31} \\
$\isotope[85][]{Se}(\alpha,n)$ & Sr/Zr, Y/Zr, Mo/Zr, Nb/Zr   & 0.23-0.38  &  26   \\
$\isotope[85][]{Br}(\alpha,n)$ & \textit{Sr/Zr}, Y/Zr, Nb/Zr, \textit{Ru/Zr, Pd/Zr} & 0.22-0.69, \textit{0.23-0.69}  & \textit{2}, 4, \textit{6}, 7, 8, \textit{10}, 20, 22   \\
~ & ~ & ~ & 25, \textit{28, 29, 30, 31} \\
$\mathit{^{86}Br(\alpha,n)}$ & \textit{Sr/Zr, Y/Zr, Nb/Zr}   & \textit{0.21-0.32}  &  \textit{6, 10, 28, 29, 31}   \\
$\isotope[86][]{Kr}(\alpha,n)$ & Sr/Zr, Y/Zr, Mo/Zr  & 0.28-0.69, \textit{0.21-0.78}  & \textit{1, 2}, 4, 7, 8, \textit{10, 17}, 20, 25, \textit{26, 31}    \\
$\isotope[87][]{Kr}(\alpha,n)$ & Sr/Zr, Y/Zr, Nb/Zr  & 0.20-0.39, \textit{0.21-0.30}  & 4, 7, 8, \textit{12, 16}, 20, 25, 30  \\
$\isotope[88][]{Kr}(\alpha,n)$ & Sr/Zr, Y/Zr, \textit{Nb/Zr}, Mo/Zr, Ru/Zr, Nb/Zr, Pd/Zr   & 0.21-0.74, \textit{0.32-0.40}  & \textit{2}, 4, \textit{6}, 7, 8, 10, 17, 20, 22  \\
~ & ~ & ~ & 25, 26, 28, 29, 30, \textit{31} \\
$\isotope[89][]{Kr}(\alpha,n)$ & \textit{Sr/Zr}, Y/Zr, \textit{Nb/Zr}, Mo/Zr, Ru/Zr, Pd/Zr  & 0.22-0.46, \textit{0.23-0.28}  &  \textit{2, 6} 10, 17, 22, 26, 29, 30, 31  \\
$\isotope[90][]{Kr}(\alpha,n)$ & \textit{Sr/Zr}, Y/Zr, Nb/Zr, Mo/Zr, Ru/Zr   & 0.26-0.53, \textit{0.21-0.22} & \textit{2, 6}, 10, 26, 29, 31  \\
$\mathit{^{87}Rb(\alpha,n)}$ & \textit{Sr/Zr, Y/Zr, Nb/Zr}   & \textit{0.37-0.66}  &  \textit{12, 13, 15, 16}   \\
$\mathit{^{88}Rb(\alpha,n)}$ & \textit{Sr/Zr, Y/Zr, Nb/Zr}   & \textit{0.35-0.40}  &  \textit{12, 13, 16}   \\
$\isotope[89][]{Rb}(\alpha,n)$ & Y/Zr, Nb/Zr, Mo/Zr, \textit{Ru/Zr, Pd/Zr}   & 0.21-0.55, \textit{0.22-0.63}  & 4, 7, 8, \textit{12, 13, 16}, 20, 25, 30   \\
$\isotope[90][]{Rb}(\alpha,n)$ & Mo/Zr & 0.24, \textit{0.26}  & \textit{7, 22, 29}, 30    \\
$\isotope[91][]{Rb}(\alpha,n)$ & \textit{Nb/Zr}, Mo/Zr & 0.27, \textit{0.20-0.38}  & \textit{6, 7, 22, 29}, 30, \textit{31}   \\
$\mathit{^{88}Sr(\alpha,n)}$ & \textit{Nb/Zr, Mo/Zr}   & \textit{0.78-0.85}  &  \textit{15}   \\
$\mathit{^{89}Sr(\alpha,n)}$ & \textit{Nb/Zr, Mo/Zr}   & \textit{0.20-0.24}  &  \textit{12, 13, 15}   \\
$\isotope[90][]{Sr}(\alpha,n)$ & Mo/Zr   & 0.28-0.33, \textit{0.75-0.79}  & 4, \textit{13, 16}, 20   \\
$\isotope[91][]{Sr}(\alpha,n)$ & Mo/Zr, \textit{Ru/Zr}   & 0.21-0.33, \textit{0.21-0.33}  & 4, \textit{7}, 8, \textit{12, 13, 16}, 20, 25   \\
$\isotope[92][]{Sr}(\alpha,n)$ & \textit{Sr/Zr}, Mo/Zr, \textit{Ru/Zr}   & 0.31-0.49, \textit{0.27-0.55}  &  4, 7, 8, \textit{16}, 20, \textit{22}, 25, \textit{28}, 30  \\
$\isotope[93][]{Sr}(\alpha,n)$ & \textit{Sr/Zr}, Mo/Zr, Ru/Zr, Pd/Zr  & 0.20-0.36, \textit{0.24-0.41}   &  \textit{6}, 7, 8, 10, \textit{22}, 25, \textit{28}, 29, 30 , 31 \\
$\isotope[94][]{Sr}(\alpha,n)$ &  \textit{Sr/Zr}, Mo/Zr, Ru/Zr, Pd/Zr  & 0.34-0.66, \textit{0.28-0.66}  & \textit{6, 7}, 10, \textit{22}, 26, \textit{28}, 29, 30, 31   \\
$\mathit{^{94}Y(\alpha,n)}$ & \textit{Y/Zr, Ru/Zr, Pd/Zr}   & \textit{0.20-0.32}  &  \textit{7, 20, 22, 25, 30}   \\
$\mathit{^{95}Y(\alpha,n)}$ & \textit{Pd/Zr}   & \textit{0.23-0.30}  &  \textit{6, 7, 22, 30}   \\
$\isotope[96][]{Y}(\alpha,n)$ & \textit{Ru/Zr}, Pd/Zr   & 0.31, \textit{0.23-0.40}  & \textit{4, 6, 7, 22}, 29, \textit{30}   \\
$\mathit{^{96}Zr(\alpha,n)}$ & \textit{Ru/Zr, Pd/Zr}   & \textit{0.24-0.74}  &  \textit{4, 7, 8, 20, 22, 25, 30}   \\
$\mathit{^{97}Zr(\alpha,n)}$ & \textit{Ru/Zr, Pd/Zr}   & \textit{0.23-0.27}  &  \textit{8, 25, 30}   \\
$\mathit{^{98}Zr(\alpha,n)}$ & \textit{Ru/Zr, Pd/Zr}   & \textit{0.21-0.29}  &  \textit{8, 22, 30}   \\
\enddata
\end{deluxetable*}

\section{Conclusions and Discussion} 
\label{sec:conclusions}

We performed a new study about the impact of the $(\alpha, n)$ reaction rates for the production of the lighter heavy elements via the weak $r$--process, using the Atomki-V2 $\alpha$OMP, and thermodynamical trajectories from~\cite{bliss2018survey}, which cover a broad range of astrophysical conditions. The $(\alpha, xn)$ reactions rates based on the successful Atomki-V2 $\alpha$OMP lead to a smaller uncertainty in the production of the lighter heavy elements (Z = 38--47). However there are still uncertain reactions that need to be studied experimentally. In particular, as we move away from stability to more neutron--rich species, these uncertainties may be larger than the predicted from current theoretical models.

It is clear from our discussion in Section~\ref{sec:compilation} that elemental observations in the first $r$--process peak of metal--poor stars are scarce. We expect that the next generation of Earth-- and space--based telescopes, such as the 4-metre multi-object spectroscopic telescope (4MOST)~\citep{de20124most}, WEAVE at the William Herschel Telescope~\citep{dalton2016final} and VLT's CUBES~\citep{genoni2022cubes, hansen2022heavy}, along with dedicated surveys, will provide more observations of metal--poor stars that will help constrain our current models and provide better predictions of the weak $r$--process. Data from more elements between strontium and silver of metal--poor stars that have already been observed in the past are also valuable.

There is a strong dependence of weak $r$--process nucleosynthesis to the entropy per baryon, expansion timescale and electron fraction of the wind ejecta. In Section~\ref{sec:conditions} we discussed which conditions can reproduce the observed elemental ratios. Our results suggest that there seems to be a relationship between the conditions of the wind and the production of Z= 38--46 elements, but more observational data are crucial to compare our models to. We should also note that even though there are Monte Carlo trajectories that failed to reproduce elemental abundance ratios in Figure~\ref{fig:6}, they should not be completely dismissed from further analysis. In explosive astrophysical environments, the neutrino--driven ejected material is described by a distribution of Y$_e$, entropy, and expansion timescale, and thus a combination of multiple such conditions could be able to reproduce the observed abundance ratios.

In this work, we identified 35 $(\alpha, n)$ reactions that impact elemental abundance ratios of Z = 38--47 elements, for different astrophysical conditions. Experimentally determining all the above cross sections and reaction rates will provide better constrains in the model parameters of the weak \textit{r}--process and also help nuclear theorists to better understand the peculiarities of the $\alpha$OMP in the intermediate nuclear mass regime.

\begin{figure}[hbpt!]
    \centering
    \includegraphics[width=.5\textwidth]{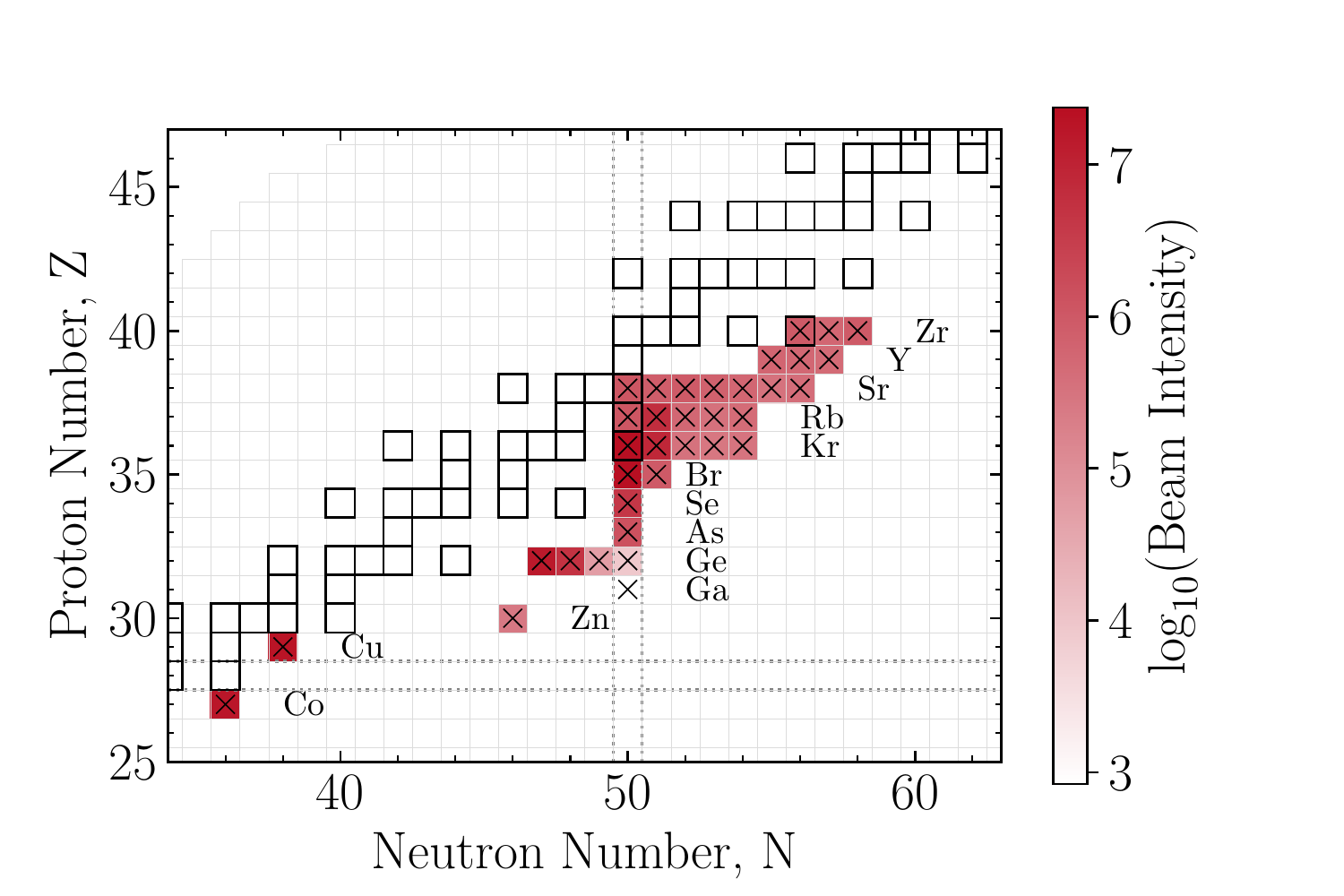}
    \caption{Target nuclei which $(\alpha,n)$ reactions affect elemental abundance ratios in different astrophysical conditions indicated with an x--mark. The color-code corresponds to estimated intensities of reaccelerated beams at the second FRIB PAC~\citep{frib-rates}.}
\label{fig:8} 
\end{figure}

It is evident from Figure~\ref{fig:8} that most of the reactions we identified as important in our study are located beyond or at the $N= 50$ shell closure nuclei, something that has been also noted in~\cite{bliss2020nuclear}. As in the main $r$--process, the $(n, \gamma) - (\gamma, n)$ equilibrium leads to an accumulation of material in the neutron shell closure nuclei which act as waiting points. $(\alpha, n)$ reactions on these nuclei help matter to move to heavier masses, and thus they are important for the weak $r$--process. Additionally, Table~\ref{tab:reactions_ratios} shows that the isotopic chains of krypton, rubidium, strontium and zirconium provide the majority of the identified $(\alpha, n)$ reactions.

It is important to emphasize that the $(\alpha,n)$ reactions we highlight in this study include nuclear species that are readily available at sufficient intensities in the current and the next generation of radioactive beam facilities, such as the Facility for Rare Isotope Beams (FRIB) -- see Figure~\ref{fig:8}, FAIR at GSI, the Californium Rare Isotope Breeder Upgrade (CARIBU) at ATLAS and ISAC/ARIEL at TRIUMF. We believe that it is a great opportunity for the experimental nuclear astrophysics community to directly measure the cross sections of the $(\alpha,n)$ reactions we identified in the present study and constrain the production of the lighter heavy elements of the first $r$--process peak.

Our work shows the potential of combining astronomical observations, hydrodynamic simulations and both theoretical and experimental nuclear physics to  understand the origin of the heavy elements in the cosmos.

\begin{acknowledgements}
This work was supported by the Deutsche Forschungsgemeinschaft (DFG, German Research Foundation)—Project No. 279384907—SFB 1245, the European Research Council Grant No. 677912 EUROPIUM, and  the State of Hesse within the Research Cluster ELEMENTS (Project ID 500/10.006). The network calculations were performed in the GSI Virgo HPC cluster. F.M. and H.S. are supported by the U.S. National Science Foundation under award numbers PHY-1913554 and PHY 14-30152 (JINA-CEE). P.M. acknowledges support from the National Research Development and Innovation Office (NKFIH), Budapest, Hungary (K134197)
\end{acknowledgements}

\textit{Software:} \texttt{h5py}~\citep{collette_python_hdf5_2014}, \texttt{IPython}~\citep{PER-GRA:2007}, \texttt{Jupyter}~\citep{Kluyver2016jupyter},
\texttt{matplotlib}~\citep{Hunter:2007}, \texttt{numpy}~\citep{harris2020array},
\texttt{pandas}~\citep{reback2020pandas}, Scientific colour maps~\citep{crameri_fabio_2021_5501399}, \texttt{seaborn}~\citep{michael_waskom_2017_883859}, \texttt{WinNet}~\citep{Winteler:2011, Winteler:2012}.


\bibliography{bibliography}{}
\bibliographystyle{aasjournal}



\appendix

\section{The $(\alpha, n)$ reaction rates that affect elemental abundances}

In Table~\ref{tab:reactions} we present the list of 37 $(\alpha, n)$ reactions
which affect \textit{elemental abundances} by more than a factor of 2 in the 2$\sigma$ distribution and also have a Spearman's coefficient $|\mathrm{r_\mathrm{corr}}|\geq0.20$. The relevant discussion can be found in Section~\ref{sec:reactions}. Note that in the work of~\cite{bliss2020nuclear}, an elemental variation of a factor of 5 was chosen as the threshold to identify a $(\alpha,n)$ reaction as important. Given that we varied
all the $(\alpha,xn)$ reaction rates by a factor of 3, compared to a factor of 10 of~\cite{bliss2020nuclear}, we assume that a lower threshold is justified. We should also point out that even when we were increasing this lower limit to factors of 4 or 5, the list of important reactions was not significantly affected.

Comparing this list of reactions with the one of~\cite{bliss2020nuclear}, we observe a very good agreement in target nuclei with $36 < Z < 41$. Our calculations were not impacted significantly by lighter mass reactions, such as $\isotope[59, 68][]{Fe}$ or $\isotope[74, 76][]{Ni}$. There are only 6 $(\alpha, n)$ reactions out of the 37 that were not included in the work of~\cite{bliss2020nuclear}, namely $\isotope[79][]{Ga}, \isotope[86][]{Br}, \isotope[91][]{Rb}, \isotope[96][]{Y}, \isotope[100, 102][]{Mo}$. In both works, the $(\alpha, n)$ reactions along the isotopic lines of krypton (Z= 36), strontium (Z= 38) and zirconium (Z= 40) are very important since the respective $\beta^-$ decays feed the first $r$--process peak elements.

\startlongtable
\begin{deluxetable}{ccccc}
\tablecaption{Element (Z) and wind trajectories for which the Spearman's coefficient is $|\mathrm{r_{corr}}|\geq0.20$ and the elemental abundance varies by more that a factor of 2 within $2\sigma$ of the abundance distribution (for elemental abundances Y$>10^{-10}$). \label{tab:reactions}}
\tablehead{\colhead{Reaction} & \colhead{Z} & \colhead{Abundance variation} & \colhead{Correlation} & \colhead{MC Tracers} \\
~ & ~ & ~ & \colhead{Coefficient, $|\mathrm{r_{corr}}|$} & ~}
\startdata
$\isotope[63][]{Co}(\alpha,n)$ &  44, 46 &  5.38-5.84  & 0.20 &  13, 20 \\
$\isotope[67][]{Cu}(\alpha,n)$ & 44, 46 & 4.69-5.38 & 0.23-0.28 & 13 \\
$\isotope[76][]{Zn}(\alpha,n)$ & 36, 41  & 2.13-3.79 & 0.22-0.23  & 1,2  \\
$\isotope[79][]{Ga}(\alpha,n)$ &  36, 41 & 2.14-2.38 & 0.33-0.38  & 1, 17  \\
$\isotope[80][]{Ge}(\alpha,n)$ & 36-39, 41, 42  & 2.14-4.81  & 0.35-0.60  & 1, 2, 17, 28  \\
$\isotope[82][]{Ge}(\alpha,n)$ & 36-39, 41, 42  &  2.14-4.34 & 0.21-0.57 & 1, 2, 17, 28  \\
$\isotope[83][]{As}(\alpha,n)$ &  36-39 & 2.93-4.97  & 0.70-0.79 &  2, 17, 26, 28  \\
$\isotope[84][]{Se}(\alpha,n)$ & 36-39, 41, 42, 44, 45  & 2.01-9.19  & 0.23-0.97 & 1, 2, 6, 7, 10, 17, 20, 22, 26, 28, 29, 30, 31 \\
$\isotope[85][]{Se}(\alpha,n)$ & 36-39, 41, 44, 45  & 2.01-8.08 & 0.22-0.36 & 1, 2, 6, 10, 17, 26, 28, 29, 31   \\
$\isotope[85][]{Br}(\alpha,n)$ & 37-39, 42, 45 & 2.09-9.19 & 0.21-0.89 & 6, 7, 10, 20, 22, 29, 30 \\
$\isotope[86][]{Br}(\alpha,n)$ &  37-39, 41 & 2.09-3.12 & 0.21-0.37 &   2, 6, 10, 22, 29, 31  \\
$\isotope[86][]{Kr}(\alpha,n)$ &  38, 40-42, 44-46 & 2.07-5.76  & 0.33-0.70  & 4, 7, 13, 20, 25 \\
$\isotope[87][]{Kr}(\alpha,n)$ & 38-42  & 2.07-2.47 & 0.20-0.38  & 4, 7, 20, 30  \\
$\isotope[88][]{Kr}(\alpha,n)$ & 38-42, 44, 45 & 2.09-5.99  & 0.21-0.73  & 2, 4, 6, 7, 10, 20, 22, 25, 26, 28, 29, 30, 31 \\
$\isotope[89][]{Kr}(\alpha,n)$ & 36, 38, 39, 42, 44, 45  & 2.07-4.18 & 0.21-0.44  & 2, 6, 26, 29, 30, 31   \\
$\isotope[90][]{Kr}(\alpha,n)$ & 36, 38, 39, 41, 42, 45  &  2.07-4.39 & 0.21-0.51 & 2, 6, 10, 26, 29, 31  \\
$\isotope[87][]{Rb}(\alpha,n)$ & 41, 42, 44  & 2.09-3.12 & 0.31-0.58  & 13, 15  \\
$\isotope[89][]{Rb}(\alpha,n)$ & 41, 42, 44-46 & 2.43-9.19 & 0.21-0.50 & 4, 7, 8, 20, 22, 25, 30, 31 \\
$\isotope[91][]{Rb}(\alpha,n)$ & 42, 45   & 2.10-2.72 & 0.24-0.54  & 6, 7, 22, 29, 30, 31   \\
$\isotope[88][]{Sr}(\alpha,n)$ & 41, 42, 44  & 2.09-3.06 &  0.30-0.44 & 15   \\
$\isotope[89][]{Sr}(\alpha,n)$ & 42  & 2.66-3.14 &  0.21-0.22  & 12, 13  \\
$\isotope[90][]{Sr}(\alpha,n)$ & 42, 44-46  & 2.43-5.38  & 0.20-0.77  & 4, 12, 13, 16, 20 \\
$\isotope[91][]{Sr}(\alpha,n)$ &  42, 44, 45  &  2.43-3.30 & 0.21-0.47  & 4, 7, 8, 12, 16, 20, 25   \\
$\isotope[92][]{Sr}(\alpha,n)$ &  42, 44-46 & 2.16-5.35  & 0.23-0.47   & 4, 7, 8, 16, 20, 22, 25, 28, 30, 31 \\
$\isotope[93][]{Sr}(\alpha,n)$ & 37, 42, 44, 46  & 2.10-5.35   &  0.21-0.42 & 6, 7, 8, 10, 22, 26, 28, 29 \\
$\isotope[94][]{Sr}(\alpha,n)$ & 37, 38, 42, 44-46 & 2.09-5.35  &  0.23-0.77  & 6, 7, 10, 22, 26, 28, 29, 30, 31 \\
$\isotope[94][]{Y}(\alpha,n)$ & 44-46  & 3.49-9.19  & 0.21-0.58  & 4, 7, 8, 20, 22, 25, 30  \\
$\isotope[95][]{Y}(\alpha,n)$ & 45, 46  & 3.25-5.99  & 0.26-0.38  & 6, 7, 8, 22, 25, 30  \\
$\isotope[96][]{Y}(\alpha,n)$ & 45, 46  & 3.25-5.99  & 0.23-0.50  & 6, 7, 8, 22, 25, 29, 30, 31  \\
$\isotope[94][]{Zr}(\alpha,n)$ & 44-46 &  3.21-5.38 & 0.21-0.40  & 12, 13, 15 \\
$\isotope[95][]{Zr}(\alpha,n)$ & 44-46 & 3.21-5.38 & 0.22-0.32 & 4, 12, 13, 16 \\
$\isotope[96][]{Zr}(\alpha,n)$ & 44-46 & 2.97-5.87 &  0.24-0.77 & , 7, 8, 12, 13, 16, 20, 22, 25 \\
$\isotope[97][]{Zr}(\alpha,n)$ & 44, 46 & 3.96-6.99  & 0.22-0.29  & 8, 25, 30 \\
$\isotope[98][]{Zr}(\alpha,n)$ & 44, 46 & 3.96-6.99  & 0.21-0.25  & 8, 22, 30 \\
$\isotope[97][]{Nb}(\alpha,n)$ & 45, 46 & 3.42-4.69  & 0.33-0.56  & 12, 13, 16 \\
$\isotope[100][]{Mo}(\alpha,n)$ & 46 &  3.42-4.69 & 0.25-0.32  & 12, 13 \\
$\isotope[102][]{Mo}(\alpha,n)$ & 46 & 3.42-4.48  & 0.21-0.28  & 4, 12 \\
\enddata
\end{deluxetable}

\end{document}